\documentclass[twocolumn,letterpaper,fleqn]{ModDetSymp}
\usepackage{fix2col}
\usepackage{multicol}
\usepackage{ifthen}
\usepackage{graphicx}
\usepackage{xcolor}
\usepackage{amsmath}
\usepackage{amsfonts}
\usepackage{lipsum}
\usepackage{hyperref}

\newcommand{\T}[1]{\mathbf{#1}}
\title{Efforts in Modeling the Mechanics and Chemistry of Energetic Materials Across Scales}
\author{Paul Lafourcade$^{\dag,\ddag}$, Nicolas Bruzy$^{\dag,\ddag}$, Paul Bouteiller$^{\dag,\ddag}$, Jean-Bernard Maillet$^{\dag,\ddag}$, Christophe Denoual$^{\dag,\ddag}$}

\affiliation{$^{\dag}$CEA, DAM, DIF, F-91297 Arpajon, France \\ $^{\ddag}$Université Paris-Saclay, LMCE, F-91680 Bruyères-le-Châtel, France}

\begin{document}


\twocolumn[
\setlength{\fboxrule}{0.5pt}
\begin{@twocolumnfalse}
\maketitle
\begin{center}
  \parbox{5in}{\textbf{Abstract. }%
Recent developments dedicated to the building of multiscale mechanical and chemical constitutive laws for energetic molecular crystals are presented and discussed. In particular, various tools have been specifically incorporated in molecular dynamics codes to facilitate the subsequent information transfer to the continuum, i.e. finite elements simulation codes. Atomistic simulations have been augmented with the capability to follow specific deformation paths as well as local Lagrangian mechanical metrics, enabling the computation of materials flow stress surface. This mechanistic library allowed the construction of a comprehensive non-linear hyperelastic continuum model including crystal plasticity and twinning for TATB. Besides, recent advances in analyzing reactive molecular dynamics simulations with unsupervised learning algorithms has enabled the identification and calibration of chemical decomposition kinetics for RDX and TATB single crystal. In the present work, the procedure is applied to $\beta$-HMX and extended with the calibration of a multi-components equation of state. These two ingredients are implemented in a finite-element code in order to model the shock-to-detonation transition at the mesoscale level and to study dimensionality effects in quasi-static hotspots. Finally, these dedicated efforts towards a comprehensive multiscale modeling of explosives has also given rise to the need for new prospective experiments, discussed throughout the paper.


\sectionline}
\end{center}
\end{@twocolumnfalse}
]

\section{1. Introduction}
\label{sec:intro}
Over the years, significant advancements have been made in the field of multiscale modeling, particularly concerning the mechanical and chemical properties of energetic materials, including high-energy compounds such as RDX, TATB, and HMX which are critical in applications ranging from propulsion systems to explosive devices. Understanding and predicting their behavior under various conditions is essential for both safety and performance optimization.

The present work focuses on the integration of molecular dynamics (MD) simulations and finite element (FE) methods to develop comprehensive constitutive models for these materials. By bridging the gap between discrete atomistic simulations and continuum-scale modeling, the present work aims to provide a more accurate and predictive understanding of energetic materials' response to mechanical and thermal stimuli.

Key to this approach is the incorporation of specific deformation paths and local mechanical metrics within MD simulations, allowing for the detailed characterization of materials' flow stress surfaces and deformation mechanisms. Additionally, recent advances in using unsupervised learning algorithms to analyze reactive MD simulations have facilitated the identification and calibration of chemical decomposition kinetics. These methodologies have been applied to RDX and TATB single crystals, and extended to HMX, culminating in the development of a multi-component equation of state (EOS).

The integration of these detailed atomic-scale insights into mesoscale FE models enables the simulation of complex phenomena such as the shock-to-detonation (STD) transition and the formation of quasi-static hotspots. These multiscale simulations not only enhance our theoretical understanding but also inform the design of new experimental setups to validate and further refine the models. Overall, this work represents a significant step forward in the multiscale modeling of energetic materials, offering new tools and approaches for the scientific community.

\section{2. Facilitating the communication between discrete and continuum scales}
\label{sec:microtomeso}
Molecular dynamics simulations represent a very powerful method that allow to compute a wide range of thermodynamic properties of materials at different conditions.

\subsection{2.1. Prescribed deformation paths in atomistic simulations}
\label{subsec:md_defpath}
When studying the mechanical response of materials upon deformation at the microscopic scale, it can be useful to deform materials in any direction in order to explore their response to a mechanical loading along any arbitrary axis. Besides, as experiments are usually conducted at constant strain rate, allowing the simulation cell to evolve dynamically appears more suited for direct comparisons. 

Two approaches can be adopted : the first one consists in orienting the material w.r.t. a fixed loading direction, e.g. the z-axis for example. \texttt{LAMMPS}~\cite{thompson_cpc_2022} has some limitations concerning triclinic cells and it is a requirement that the $\T{a}$ vector is parallel to the x-axis, the $\T{b}$ vector has to belong to the (a,b) plane. Such conditions imply that various deformations cannot be performed using \texttt{LAMMPS}, in particular the ones that shifts the $\T{a}$ and $\T{b}$ vectors from the x-axis and (x,y) plane, respectively. Ensuring that the \texttt{LAMMPS} convention is always respected implies some limitations on the deformations that can be applied to the simulation domain, especially in the presence of low symmetry molecular crystals commonly found in energetic materials such as TATB, HMX and RDX among others.

To overcome this limitation, a methodology to perform 3D periodic MD simulations with controlled large deformations was specifically designed in the \texttt{exaNBody} code~\cite{carrard_2024}. It allows to perform uniaxial, shear, triaxial or non-proportional deformations by providing analytical expressions of a time-dependent deformation gradient tensor. In addition, the core design of \texttt{exaNBody} does not require any conditions on the periodic vectors of the simulation cell, leaving any possibility for applied deformations, including large rotations. The initial MD simulation frame tensor $\T{H_0}$ is then defined as:
\begin{equation}
    \T{H_0}=\begin{pmatrix} a_x & b_x & c_x \\ a_y & b_y & c_y \\ a_z & b_z & c_z\end{pmatrix}
\end{equation}
where $\T{a}$, $\T{b}$ and $\T{c}$ are the vectors that defined the direction of periodic boundaries of the MD simulation cell. 
An additional option, used in the present work, allows the user to inform the MD engine a list of deformation gradient tensors along with a list of physical times. This set of deformation gradient tensors is then interpolated by parts during the MD trajectory, leading to the following instantaneous frame tensor $\T{H}(t)$:
\begin{equation}
    \T{H}(t)=\T{F}(t)\T{H_0}
    \label{eq:tdepstrain}
\end{equation}
where $\T{F}(t)$ is the deformation gradient tensor in the current state. In doing so the user can design specific deformation gradient tensors, allowing to trigger various deformation mechanisms and study directional dependence of mechanical response of materials, e.g. as was previously done for TATB single crystal~\cite{lafourcade_prm_2019}. Concerning the second alternative to deform materials along specific directions, e.g. pre-generating oriented atomistic samples, different codes have been developed in recent years. Both the Generalized Crystal-Cutting Method (GCCM)~\cite{kroonblawd_cpc_2016} and the Los Alamos Crystal Cut (LCC)~\cite{negre_jpcm_2023} constitute powerful crystal builders dedicated to atomistic simulations. In particular, they help generating 3D-periodic systems for low index molecular crystals. However, when oriented samples or deformation paths are used during dynamic deformation with 3D-periodic boundary conditions, different responses can be obtained for small samples as periodic boundary conditions may lead to self-interactions of defects such as twins, phase transformations or dislocations.  

With the emergence of experimental mechanics at the nanoscale, the developments of such tools is of prime importance as they allow for a step towards qualitative comparisons. Traction-compression experiments at the nanoscale are becoming routine when dealing with metals or ceramics , although very few results are available concerning energetic molecular crystals. Experiments such as in-situ SEM nanopillar compression~\cite{guruprasad_asm_2023} or traction-compression on single crystal samples~\cite{franciosi_ijp_2015} provide some insight into the deformation mechanisms of materials such as amorphization, twinning deformation or plastic slip activity. However, as example for EBSD scanning, the irradiation required for the analysis of crystal orientation deteriorates the surface state of the material, preventing any diagnostic. It is therefore crucial to continue developing methods to fill the gap in experimental facilities for the small-scale (quasistatic) mechanics of energetic materials.

\subsection{2.2. Discovery of deformation mechanisms using the flow stress surface}
\label{subsec:flowsurface}
The entire TATB single crystal flow surface was previously obtained through classical MD simulations~\cite{lafourcade_jpcc_2018} at ambient conditions using the deformation pathways formalism explained previously. It represents the orientation dependence of the macroscopic stress at onset of the first deformation mechanism, gathering all the elementary deformation processes occurring at the microscopic scale for this material. The flow surface was obtained under planar strain conditions and directional pure shear. Following Eq.\ref{eq:tdepstrain}, the applied deformation gradient tensor had the following form:
\begin{equation}
    \T{F}(t) = \T{I} + \alpha(t) \T{m} \otimes \T{m} + \beta(t) \T{p} \otimes \T{p},
\label{eq:defpath}    
\end{equation}
with $\alpha$ and $\beta$ two time dependent functions, $\otimes$ is the dyadic product and $\T{I}$ is the second-order identity tensor. $\T{m}$ and $\T{p}$ were chosen orthogonal and act as a compression and traction direction, respectively. They define a plane chosen to be perpendicular to the (x,y) plane, i.e. rotating only around the z-axis, parallel to the symmetry axis of TATB if one assumes a transverse isotropy symmetry. $\alpha(t)$ and $\beta(t)$ were constructed such that deformations are performed at iso-volume and at a constant strain-rate $\dot{\varepsilon}$=10$^8$ s$^{-1}$. A total of 84 MD trajectories were then performed to compute TATB single crystal directional response to pure shear, leading to the flow surface displayed in Figure~\ref{fig:MD_vs_meso_flow}a. Each point on this surface corresponds to the von Mises stress $\sigma^\mathrm{vM}$ at nucleation of the first defect and/or at the activation of a corresponding deformation mechanism. Three distinct zones emerge from the microscopic flow surface:
\begin{itemize}
\item Superior and inferior lobes are associated with transverse dislocation-mediated plasticity in slip systems listed in \cite{lafourcade_jpcc_2018}, with large values of $\sigma^\mathrm{vM}$,
\item Basal gliding under pure shear of the $\T{(001)}$ plane leads to the area with the lowest values of $\sigma^\mathrm{vM}$,
\item Finally, a twinning-buckling elastic instability, nucleated upon compression within the molecular layers, i.e. directions lying in the $\T{(001)}$ basal plane is related to the \emph{donut}-like shape..
\end{itemize}
The microscopic flow surface of TATB is extremely rich as it synthesizes most (if not all) deformation mechanisms of the single crystal. It proves that TATB single crystal possesses an anisotropy in terms of activated deformation mechanisms, in addition to its well known structural and elastic anisotropy. 
While the present paper focuses on TATB mechanical behavior, the microscopic flow surface can be of interest for a wide range of materials with different crystal structures and help understanding their mechanical response to different loading types and directions. Moreover, the microscopic flow surface can also be computed for different types of loading such as traction/compression, in order to investigate the symmetry of the response of the material. Finally, it may enable the discovery of new deformation mechanisms, as the buckling instability for TATB. Informing constitutive laws at the mesoscale with microscopic data usually requires that the continuum model reproduces features predicted with atomistic methods. Beyond EOS or elastic behaviour, we believe that the microscopic flow surface is a strong constraint for the development and validation of mesoscopic laws.

TATB microscopic flow surface has been involved in the validation of a multiscale constitutive law for its mechanical behavior and especially in the calibration of slip systems critical stresses. This will be discussed in Section~\textbf{2}. Future work will focus on applying this method to other HE molecular crystals and especially to $\beta$-HMX, for which a thermochemistry model will be discussed in Section~\textbf{3}.

\subsection{2.3. Lagrangian deformation measures at the atomic level}
Transformation of solids with respect to a reference state can be measured through the deformation gradient tensor $\T{F}$, defined as $\T{dx} = \T{F}\cdot\T{dX}$ transforming a material vector in the reference state $\T{dX}$ to the same material vector in the current state $\T{dx}$. 
This second order tensor contains all the information relative to the modification of the local environment except translation. The measure of the deformation gradient tensor $\T{F}_\alpha$ at the atomic-level was implemented in the ExaStamp MD code based on previous work~\cite{lafourcade_jpcc_2018} and is also available in Ovito~\cite{ovito}. This tensor is the gradient of the best interpolation field of neighboring atoms displacements, relatively to a central atom $\alpha$:
\begin{multline}
\T{F}_\alpha =  \left( \sum_{\beta \in \mathcal{V}(\alpha)} \omega(r_{\alpha\beta}) \T{x}^{\alpha\beta} \otimes \T{X}^{\alpha\beta} \right) \cdot \\ \left( \sum_{\beta \in \mathcal{V}(\alpha)} \omega(r_{\alpha\beta}) \T{X}^{\alpha\beta} \otimes \T{X}^{\alpha\beta} \right)^{-1}
\end{multline}
where $\beta$ denotes the neighboring atoms and $\T{X}^{\alpha\beta}$, $\T{x}^{\alpha\beta}$ refer to the vectors between atoms $\alpha$ and $\beta$ in the reference and current states respectively. 
A weight function $\omega$ is assigned to each neighbor to favor the contribution of first and second neighbors in the presence of large deformation.

Accessing the information at the atomic scale using such metrics is of primary importance when performing mechanical loading.
Indeed, a direct comparison of mechanical fields such as the normof the Green-Lagrange strain tensor can be useful to validate or invalidate models.
Finally, developments of experiments in which local information can be accessed are more and more common nowadays. Deformation measures at the atomic scale can also be used to understand, interpret and validate/invalidate deformation processes extracted from experimental fields obtained with digital image correlation (DIC)~\cite{freville_rsi_2023}, electron back scattered diffraction (EBSD) or even X-ray tomography~\cite{henry_am_2024}.

\section{3. Non-linear hyperelastic model with crystal plasticity and twinning for TATB}
\label{sec:tatb}
\subsection{3.1. ExaCoddex formalism}
In the present section, an elastic-plastic-transformational mesoscale model, implemented in the exaCoddex in house code, for the  mechanical response of TATB is presented. The constitutive law emerges from the gathering of different studies, and includes a non-linear hyperelasticity \cite{lafourcade_prm_2019} coupled with an isothermal EOS, a phase-field by reaction pathways (PFRP) formalism to account for twinning~\cite{denoual_prl_2010} as well as a crystal plasticity framework~\cite{bruzy_jmps_2022,lafourcade_jap_2024}. This mesoscale model is entirely informed by atomistic simulations which is at the core of the present work. It is integrated in an element-free Galerkin (EFG) least-squares formulation code within a 3D Lagrangian formalism and explicit time-integration.

Upon the Lagrangian formalism, plastic activity is defined in the isocline configuration and prior to any twinning or elastic deformation, leading to the following multiplicative decomposition:
\begin{equation}
    \T{F}=\T{F}_e \cdot \T{F}_t \cdot \T{F}_p
\end{equation}
with $\T{F}_e$, $\T{F}_t$ and $\T{F}_p$ the elastic, transformational and plastic distortions. In the following, we rapidly present the different elements of the constitutive model, each involving different physics. 

\subsection{3.2. Elasticity and equation of states}
\begin{figure}[!h]
    \centering
    \includegraphics[width=0.75\linewidth]{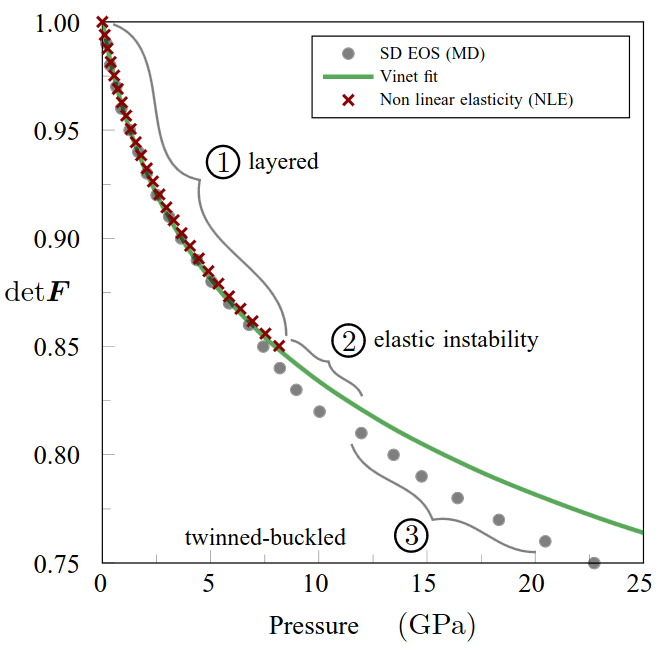}    
  \caption{Volume deformation EOS for TATB single crystal. MD, model and Vinet fit ($K_0$=17.87 GPa, $K'$=12.75) are represented as gray marks, red crosses and green line respectively.}
  \label{fig:eos_tatb}
\end{figure}

The first originality of the mesoscale model for TATB resides in its non-linear hyperelasticity formulation. The hyperelastic potential $\psi_e$ includes an evolution of stiffness w.r.t. volume deformation~\cite{lafourcade_prm_2019} and reads:
\begin{equation}
    \rho_0 \psi_e = \frac{1}{2} \T{E}_e:\mathbb{D}(\mathrm{det}\T{F}):\T{E}_e,
\end{equation}
with $\T{E}_e$ the Green-Lagrange strain tensor built on $\T{F}_e$. At this time, no temperature dependence is included in the model. The evolution of pressure $p=-\rho_0 \partial \psi_e/\partial V$ with volume $V$ is ensured by considering a volume-dependent stiffness tensor $\mathbb{D}(\mathrm{det}\T{F})=k(\mathrm{det}\T{F}) \mathbb{C}_0$ with $\mathbb{C}_0$ the elasticity at ambient conditions and $k$ a non-linear scalar function. The latter is determined starting with the Piola-Kirchhoff tensor $\T{P}$:
\begin{equation}
\begin{split}
\T{P} & =\rho_0\frac{\partial \psi_e}{\partial \T{F}} \\
& = \T{F}_e \cdot \mathbb{D}:\T{E}_e \cdot \T{F}_{in}^{-T} + \frac{1}{2}\T{E}_e:\frac{\partial \mathbb{D}}{\partial \T{F}}:\T{E}_e
\end{split}
\end{equation}
with $\T{F}_{in}=\T{F}_t\T{F}_p$ the inelastic part of the deformation. The Cauchy stress tensor $\T{\sigma}$ then reads:
\begin{equation}
\begin{split}
\T{\sigma} & = \frac{1}{J} \T{P}:\T{F}^T \\
& = \frac{1}{J} \T{F}_e \cdot \mathbb{D}:\T{E}_e \cdot \T{F}_{e}^{T} + \frac{1}{2}\T{E}_e:\mathbb{D}':\T{E}_e\cdot \T{I}
\end{split}
\end{equation}
with $J=\mathrm{det}(\T{F})$ and $\mathbb{D}'=\partial \mathbb{D}/\partial \mathrm{det}\T{F}$. Imposing that the pressure derives from an EOS:
\begin{equation}
    -\frac{1}{3}\mathrm{trace}(\T{\sigma})=P_\mathrm{EOS}(\mathrm{det}(\T{F})),
\end{equation}
allows to numerically solve the first-order differential equation on $k$ for a volume deformation $\T{F}=(V/V_0)^{1/3}\T{I}$. That process is computed once and for all at the beginning of the simulation and allows the definition of an elasticity that satisfies the material's EOS. This is illustrated in Figure~\ref{fig:eos_tatb} where the pressure from both the model and MD are compared to a Vinet EOS.

\subsection{3.3. Twinning deformation pathways}
\begin{figure}[!h]
    \centering
    \includegraphics[width=\linewidth]{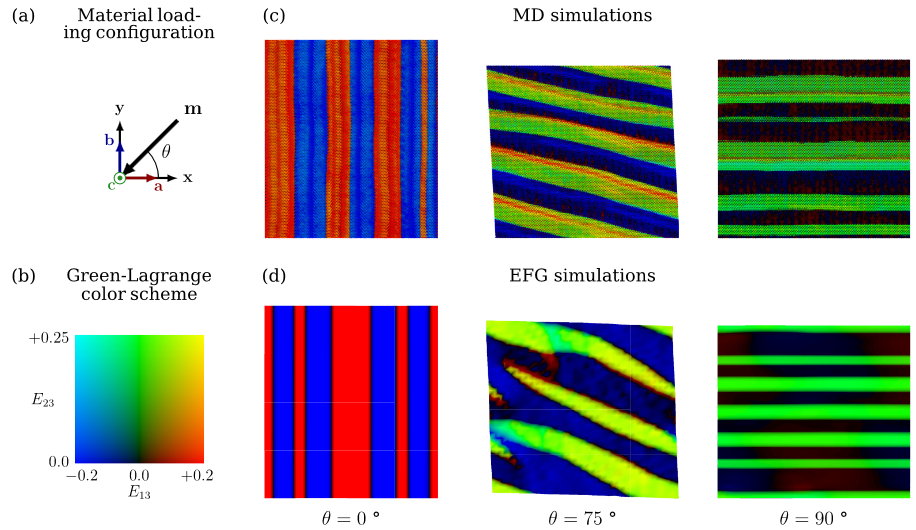}    
  \caption{a) Loading orientation. (b) Green-Lagrange color scheme. (c) Snapshots of
three MD simulations. (d) Snapshots of three mesoscale simulations.}
  \label{fig:MD_vs_meso}
\end{figure}

The second feature of the mesoscale model is a nonsymmorphic twinning deformation mechanism that was discovered through classical MD calculations~\cite{lafourcade_jpcc_2018}. Consider a centrosymmetric unit cell of lattice length $\T{a}$, $\T{b}$ and $\T{c}$, with two TATB molecules and in-plane angle $\gamma$ and out of planes angles $\alpha$ and $\beta$. A pure shear (maintaining the interplanar distance constant) that leads to new out of planes angles $\alpha'=\pi-\alpha$ and $\beta'=\pi-\beta$ allows to recover the centrosymmetric unit cell of TATB but rotated by $\pi$ around the z-axis. The inversion of $\alpha$ and $\beta$ out-of-plane angles can be obtained by applying a homogeneous shear to the TATB single crystal basal plane. 
In Figure~\ref{fig:MD_vs_meso} a comparison between the constitutive law that include both twinning and nonlinear elasticity and direct MD simulations is performed for three different loading directions. 

Depending on the loading direction, either positive/negative buckling (red/blue) or twinning (green) mechanisms is triggered. 
It was shown that in either case the buckling instability (out of plane shift of molecular layers) is responsible for the observed layered microstructure
However, the initial loading orientation fully determines the balance between the reversible buckling or twinning mechanisms. 

The phase-field by reaction pathways (PFRP) method, initially dedicated to model martensitic transformations~\cite{denoual_prl_2010}, and applied to pressure-induced phase transformation of iron~\cite{vattre_jmps_2016} and twinning of tantalum~\cite{bruzy_jmps_2022} also proves to be for modeling the twinning mechanism of TATB~\cite{lafourcade_prm_2019}. Finally, including both twinning and nonlinear elasticity in the constitutive law is sufficient to reproduce the buckling elastic instability observed in atomistic simulations.

This one-to-one validation of the constitutive law allowed for large-scale simulations of low-velocity shock loading of TATB polycrystal with sizes unreachable to MD simulations. Intragranular deformation bands were observed due to the intergranular interactions, similar to microstructural features present in optical microscopy observations~\cite{trumel_europyro_2019}. While additional observations still lack for TATB-based explosives, recent progresses have been made for getting insights into HE microstructure through X-ray tomography for example~\cite{yeager_mat_2020}. Enabling imaging of HEs using scanning or transmission electron microscopy (respectively SEM or TEM) and EBSD could really helps understand their mechanical behavior at lower scales but still represents a challenge.

\subsection{3.4. Crystal plasticity calibration using the microscopic flow stress surface}
\begin{figure}[!h]
    \centering
    \includegraphics[width=0.9\linewidth]{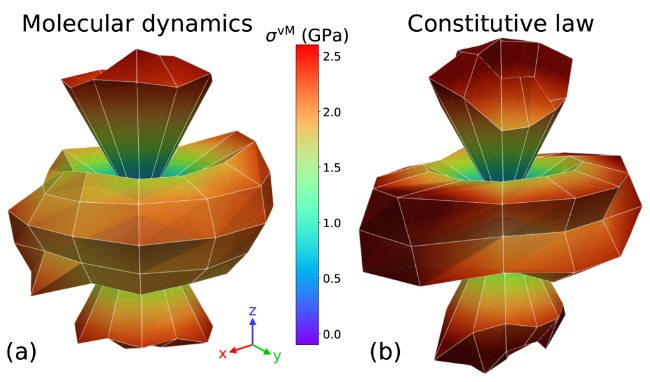}    
  \caption{Comparison between MD and mesoscopic single crystal flow surface.}
  \label{fig:MD_vs_meso_flow}
\end{figure}

Finally, the mesoscale constitutive law for TATB was recently augmented with a crystal plasticity feature. 
The various slip systems among four different nonbasal planes have been identified from MD simulations results but the calibration of the model was impeded by a lack of data. 
We developed instead an original approach~\cite{lafourcade_jap_2024} which consists in using the microscopic flow stress surface to calibrate the critical shear stresses of the identified slip systems for TATB single crystal. The microscopic flow surface computed using MD was used as the ground truth to optimize the critical shear stresses of the mesoscale model. This procedure led to critical shear stresses $\tau_c=194$ MPa and $\tau_c=960$ MPa for basal and transverse slip systems, respectively. Figure~\ref{fig:MD_vs_meso_flow} displays the comparison of both MD and mesoscale model flow stress surfaces. 
The very good agreement between the two indicates that the mesoscale model is able to reproduce the directional dependence of the deformation mechanisms activation and indeed contains all the necessary physical ingredients.

The model containing nonlinear elasticity and crystal plasticity was then used to simulate the uniaxial loading under high strain-rate of TATB single crystal, leading to different timescale of slip-systems activation and resulting microstructures~\cite{lafourcade_jap_2024}.

\section{4. Modeling the shock-to-detonation transition in HMX at the mesoscale}
\label{sec:hmx}
\subsection{4.1. Charon/FEnicS formalism}
In the present section, a mesoscale model involving the coupling between a thermochemistry model for $\beta$-HMX and a multi-components EOS is discussed. The full model is implemented our in house hydrodynamic code CHARON, leveraging the FEniCSx ~\cite{barrata2023dolfinx} open-source library, designed to address the elastoplastic finite strain problem with phase transitions using the finite element method (FEM). This multi-field problem is tackled sequentially by integrating a conventional explicit strategy with standard isoparametric elements for displacement, whereas, the evolution of the concentrations of various constituents is resolved on a cell-by-cell basis through Discontinuous Galerkin elements. The use of the Unified Form Language (UFL)~\cite{alnaes2014unified} facilitates the symbolic representation of various evolution laws, thus enabling seamless dimensional scalability (1D, 2D, 3D) without the necessity for code modifications. 

\subsection{4.2. Unsupervised learning of decomposition kinetics using reactive atomistic simulations}
The use of reactive MD simulations has allowed to gain more insights into HEs decomposition kinetics. These simulations usually employ the ReaxFF force field or its variants~\cite{van_duin_jpca_2001,wood_prb_2018}, although more elaborated reactive machine learning FF have recently been proposed~\cite{yoo_npj_2021}. Alternatively, more accurate methods such as density-functional tight binding (DFTB)~\cite{Cawkwell_jcp_2019} may also be used.

Using both reactive MD with ReaxFF and unsupervised learning methods, mesoscale reactive models for both RDX~\cite{sakano_jpca_2020} and TATB~\cite{lafourcade_jpcc_2023} have recently been obtained. 
By monitoring the time-evolution of chemical environments of CHNO systems, 280 variables are extracted at each analysis time, which represent the instantaneous chemical state of the system.
Then, using the non-negative matrix factorization (NMF) method~\cite{fevotte_nc_2011}, this dimension can be reduced to a smaller number of components, typically three or four, while ensuring the correct reproduction of the whole MD trajectories.
The optimal number of components represents a compromise between a good reproduction of the initial 280 dimensions and a reasonable dimension in the mesoscale model. This model is based on Arrhenius laws considering only irreversible first order chemical reactions. In the presence of 4 components, it reads:
\begin{equation}
\begin{split}
  \dot{C}_1 &=-C_1 Z_a \exp \big( - \frac{E_a}{RT}\big) \\
  \dot{C}_2 &= C_1 Z_a \exp \big( - \frac{E_a}{RT}\big) - C_2 Z_b \exp \big( - \frac{E_b}{RT}\big)  \\
  \dot{C}_3 &= C_2 Z_b \exp \big( - \frac{E_b}{RT}\big) - C_3 Z_c \exp \big( - \frac{E_c}{RT}\big)  \\
  \dot{C}_4 &= C_3 Z_c \exp \big( - \frac{E_c}{RT}\big)
\end{split}
\label{eq:arrhenius}
\end{equation}
with $Z_i$ and $E_i$ the kinetic prefactors and activation energies. The corresponding heats of reaction are also calibrated based on the heat balance equation along adiabatic simulations : 
\begin{equation}
\rho C_v \dot{T} = - Q_1 \dot{C}_1 - Q_2 (\dot{C}_1+\dot{C}_2) + Q_3 \dot{C}_4,
\label{eq:heat_evolved_equation}    
\end{equation}
with $\rho$ the material density, $C_v$ the classical specific heat of the reacting system, and $Q_1$, $Q_2$, $Q_3$ the heats of reaction. Finally, the calibrated model is coupled to a diffusion term and used in 1D simulations. The very good agreement obtained between continuum and MD simulations for the propagation of 1D hotspots for both RDX~\cite{sakano_jpca_2020} and TATB~\cite{lafourcade_jpcc_2023} validates this approach.
\begin{table}[!h]
\caption{4-components model kinetics parameters and heats of reaction for $\beta$-HMX.}
\centering
\begin{tabular}{c c c r}
    \hline
    Step & Z$_i$ [ps$^{-1}$] & E$_i$ [kcal/mol] & Q$_i$ [kcal]\\ 
    \hline
    1-2  & 8.981 & 13.501 & 4.5571$\times10^{5}$\\
    2-3  & 6.024 & 18.061 & 1.7639$\times10^{6}$\\
    3-4  & 1.362 & 18.329 & 3.3424$\times10^{5}$\\ 
    \hline 
\end{tabular}
\label{tab:ArrheniusHMX}
\end{table}
The exact same methodology was applied to $\beta$-HMX resulting in a 4-components thermochemistry model (1: $\beta$-HMX, 2,3: intermediates, 4: products). The optimized parameters of Eq~\ref{eq:arrhenius} and~\ref{eq:heat_evolved_equation} were calibrated on reactive MD trajectories and are reported in Table~\ref{tab:ArrheniusHMX}. 
Both temperature and concentrations evolution are reported in Figure~\ref{fig:validation_model}a-b.

\subsection{4.3. Calibration of multi-components equations of state}
In order to model the shock-to-detonation (STD) transition in $\beta$-HMX, the thermochemistry model at the mesoscale has to be coupled to a multi-components EOS. Indeed, the STD transition implies an overpressure wave that catches the inert shock wave due to the reaction of the material. 
In order to calibrate EOS on different ($\rho,T,P$) states before, during, and after the decomposition process, adiabatic decomposition simulations were performed at $0.9\rho_0$ and $0.8\rho_0$. Additionally, both the cold and 300 K isothermal EOS of $\beta$-HMX were calculated using ReaxFF in order to get a reliable EOS over a wide ($\rho$,P) interval. This entire set of points constitutes a database on which a multi-components EOS for $\beta$-HMX can be trained.
During the adiabatic decomposition, we assume, independently from time $t$, that the total instantaneous pressure can be decomposed as the sum of partial pressures from individual components:
\begin{equation}
P_{tot}(\rho,T)=\sum_{i=1}^{4} P^{eos}_i(\rho,T) \cdot C_i(\rho,T)
\end{equation}

We consider that the total pressure at early times only comes from the neat HMX contribution. Similarly, the total pressure at the end of the decomposition is given by the detonation products EOS only. These assumptions allow us to decouple the different partial EOS and the calibration process following:
\begin{itemize}
    \item[1.] The MD $\beta$-HMX cold curve is used to calibrate a 3-parameters MACAW EOS~\cite{Lozano_jap_2022} (cold part) for the 1st component (inert $\beta$-HMX),
    \item[2.] The MD data before reaction is used to calibrate an analytical and complete EOS~\cite{Lozano_jap_2023} for the 1st component (inert $\beta$-HMX),
    \item[3.] The MD data after reaction is used to calibrate an analytical and complete EOS~\cite{Lozano_jap_2023} for the 4th component (detonation products).
\end{itemize}
At this stage, two EOS namely $P^{eos}_1(\rho,T)$ and $P^{eos}_4(\rho,T)$ for both the 1st (inert) and 4th (products) components are calibrated on the MD data. Concerning the EOS of the reaction intermediates, we first consider the same EOS for each intermediate ($P^{eos}_{23}=P^{eos}_2=P^{eos}_3$). We show below that the good agreement with MD results justifies this assumption.
The calibration of the intermediates EOS $P^{eos}_{23}(\rho,T)$ is performed using the following expression:
\begin{equation}
\begin{split}
P^{eos}_{23} =( P^{MD}&-P^{eos}_1\cdot C_1\\
&-P^{eos}_4\cdot C_4 ) / (C_2+C_3)
\end{split}
\end{equation}
where each term depends on density $\rho$ and temperature $T$. The full calibration of the EOS of intermediates is performed on the entire $(\rho,T,P)$ database. 
\begin{figure}[!h]
    \centering
    \includegraphics[width=0.75\linewidth]{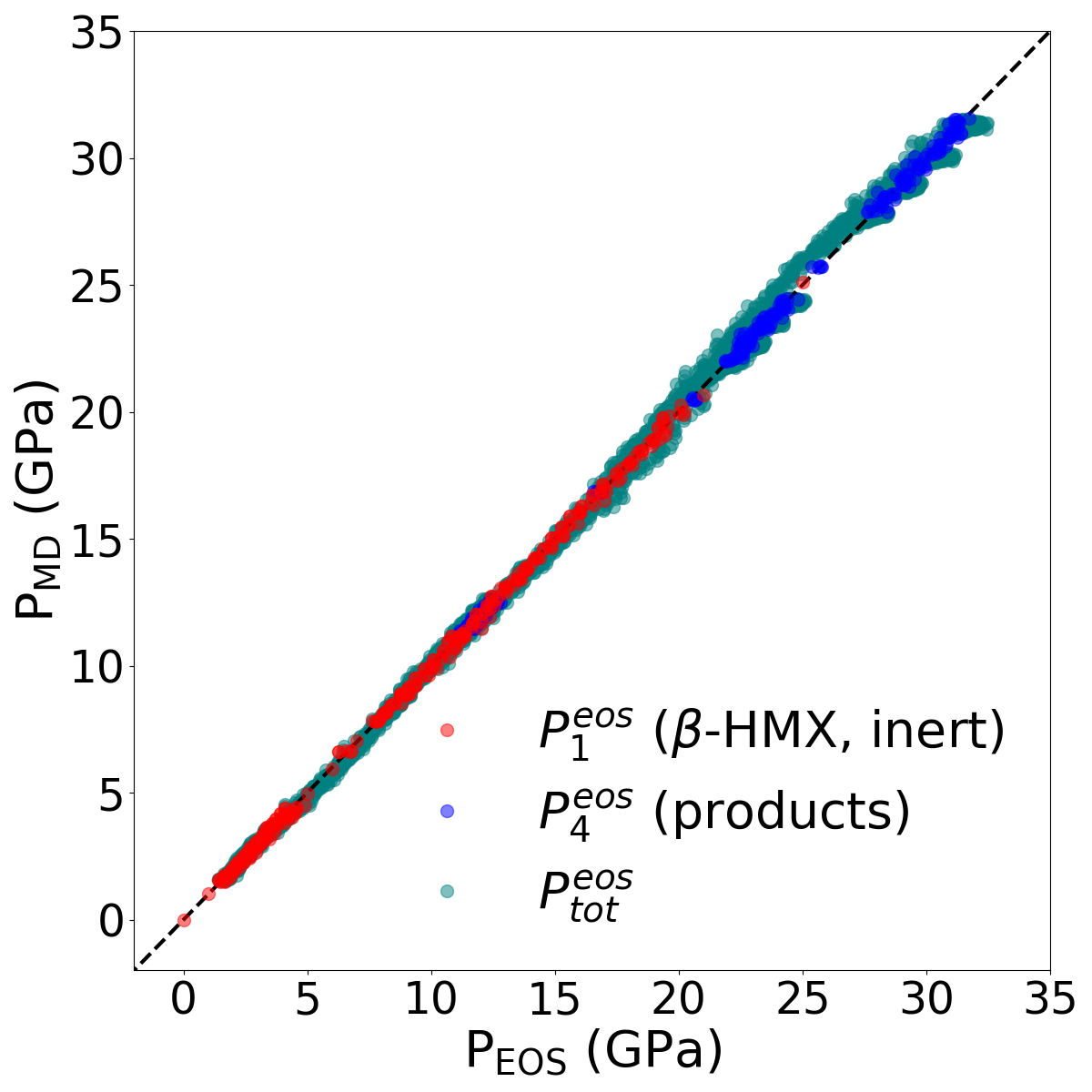}    
  \caption{Predicted pressure using the multi-components EOS vs. calculated pressure with MD calculations for $\beta$-HMX.}
  \label{fig:parity}
\end{figure}
The comparison between the multi-components EOS and MD is provided in Figure~\ref{fig:parity} using a parity plot where the pressure predicted by the analytical EOS is plotted as a function of the pressure obtained by MD. A very good agreement is found between the model and the discrete data, indicating that the calibrated EOS allows a good prediction of all components partial pressures. This multi-components EOS will be of great interest when dealing with shock-wave propagation problems and especially in the case of the shock-to-detonation (STD) transition of $\beta$-HMX. 
As a validation case, the multi-components EOS is coupled to the thermochemistry model in order to compare with MD the temperature, concentrations and pressure evolution with time under homogeneous adiabatic decomposition conditions, i.e. without any spatial diffusion term yet. Three different initial temperatures ($T_0 \in [1100, 1600, 2100]$ K) are considered and assigned to the $\beta$-HMX simulation cell at $t=0$ for both MD and mesoscale simulations. The system is then evolved under NVE conditions and the chemical decomposition inherently triggered, with more or less delay time that depends on $T_0$. 
\begin{figure*}[!t]
    \centering
    \includegraphics[width=\linewidth]{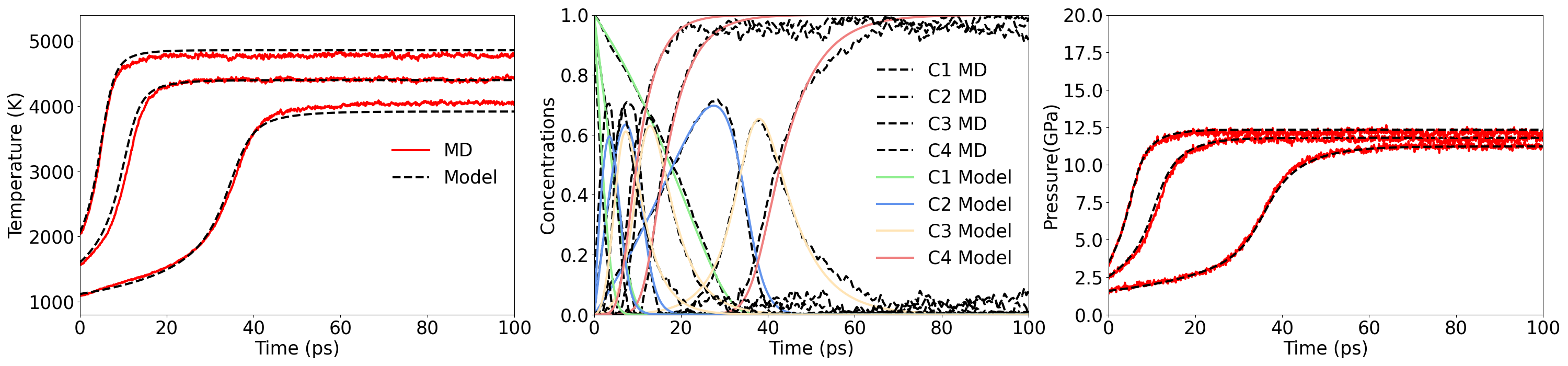}    
  \caption{Validation of the mesoscale thermochemistry model coupled with a multi-components EOS. (a) Temperature evolution for different initial temperatures. (b) Concentrations evolution. (c) Pressure evolution.}
  \label{fig:validation_model}
\end{figure*}
Temperature evolution is displayed on Figure~\ref{fig:validation_model}a where the influence of $T_0$ on the reaction kinetics is clearly visible. The mesoscale model (dashed line) is in very good agreement with the MD results (red). Figure~\ref{fig:validation_model}b reports the evolution of the four components for the three cases with striking agreement between MD and the model. Finally, during the problem resolution, both $\rho$ and $T$ are taken as input by the multi-components EOS that predicts the instantaneous pressure $P$ displayed in Figure~\ref{fig:validation_model}c. In conclusion, the full thermochemistry+EOS mesoscale model predicts the correct evolution of temperature, concentrations and pressure which proves appropriate to study hotspot criticality and STD at scales unreachable to MD simulations.

\subsection{4.4. Static dimensional effects in hotspots}
Ignition and growth of hotspots is a key feature to the developments of reliable models at the mesoscale for shock initiation of energetic materials~\cite{handley_2012}. The thermochemistry+EOS model previously developed is coupled to a diffusion equation to study hotspot criticality in $\beta$-HMX. The thermal conductivity is set to 0.497 W.m$^{-1}$.K$^{-1}$ close to experimental data~\cite{perriot_jap_2021} and a temperature-dependent specific heat used~\cite{Lozano_jap_2023}. Simulations are performed using 1D Cartesian, 1D Cylindrical and 1D Spherical meshes to study the dimensionality effects. It was carefully checked that 1D Cylindrical and Spherical meshes reproduced the exact same solutions as 2D and 3D Cartesian meshes, respectively. 
\begin{figure}[!b]
    \centering
    \includegraphics[width=0.95\linewidth]{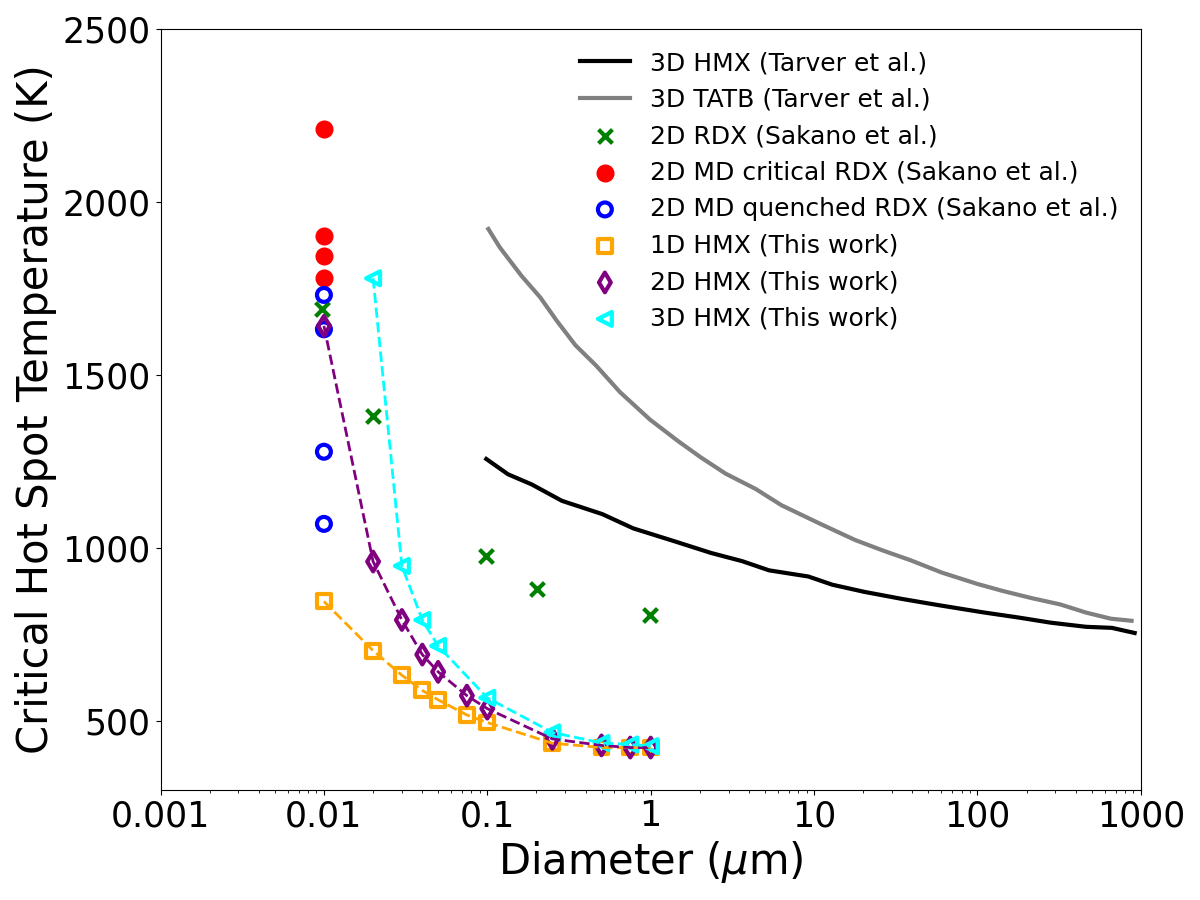}    
  \caption{Hotspot temperature vs critical size from both MD (RDX~\cite{sakano_jpca_2020}) and continuum models (RDX~\cite{sakano_jpca_2020}, TATB and HMX~\cite{tarver_1996}).}
  \label{fig:criticality}
\end{figure}
Hotspots diameters $\phi \in [10 \mathrm{nm},1 \mu\mathrm{m}]$ were considered and critical temperature identified using a dichotomy algorithm. To determine whether the hotspot went critical, a spatial domain between the initial hotspot radius and 1.2 that value was monitored. If the average products concentration in that domain surpassed 0.5 after 50 ns, the hotspot was considered as critical. A similar diagnosis was considered for RDX~\cite{sakano_jpca_2020} due to the induction time coming from the competition between diffusion and reaction processes. Results are reported in Figure~\ref{fig:criticality}. Dimensionality has an effect on hotspot critical temperature at low diameters but the differences reduce with hotspot size. The results show that our continuous model strongly underestimates the critical temperature in comparison with the HMX model from Tarver~\cite{tarver_1996} (black line) calibrated on experiments. The present results indicate that either the decomposition kinetics are too fast or the delivered energy too high, leading to an overheating of the hotspot at low temperatures that takes over the diffusion process. It was previously discussed that the rates of decomposition obtained with ReaxFF-2018 are in general faster than other variants and models calibrated on experiments~\cite{hamilton_jpca_2021}. We believe that, even if our path towards the multiscale modeling of HEs chemical decomposition looks very promising, high care must be taken concerning the identified kinetics. Future work will be dedicated to performing adiabatic decomposition using more appropriate methods. In particular, adiabatic decomposition using the ReaxFF-LG parametrization will be performed to study the impact on the final mesoscopic model, should the kinetics be very different. Another path would be to perform adiabatic decomposition using DFTB calculations, as previously done on other HEs~\cite{perriot_jap_2021}. However, such simulations are limited in size an simulated time, which might not be suitable for extracting representative kinetics for a mesoscale model.

\subsection{4.5. Shock-to-detonation transition in HMX single crystals}
\begin{figure}[!h]
    \centering
    \includegraphics[width=0.95\linewidth]{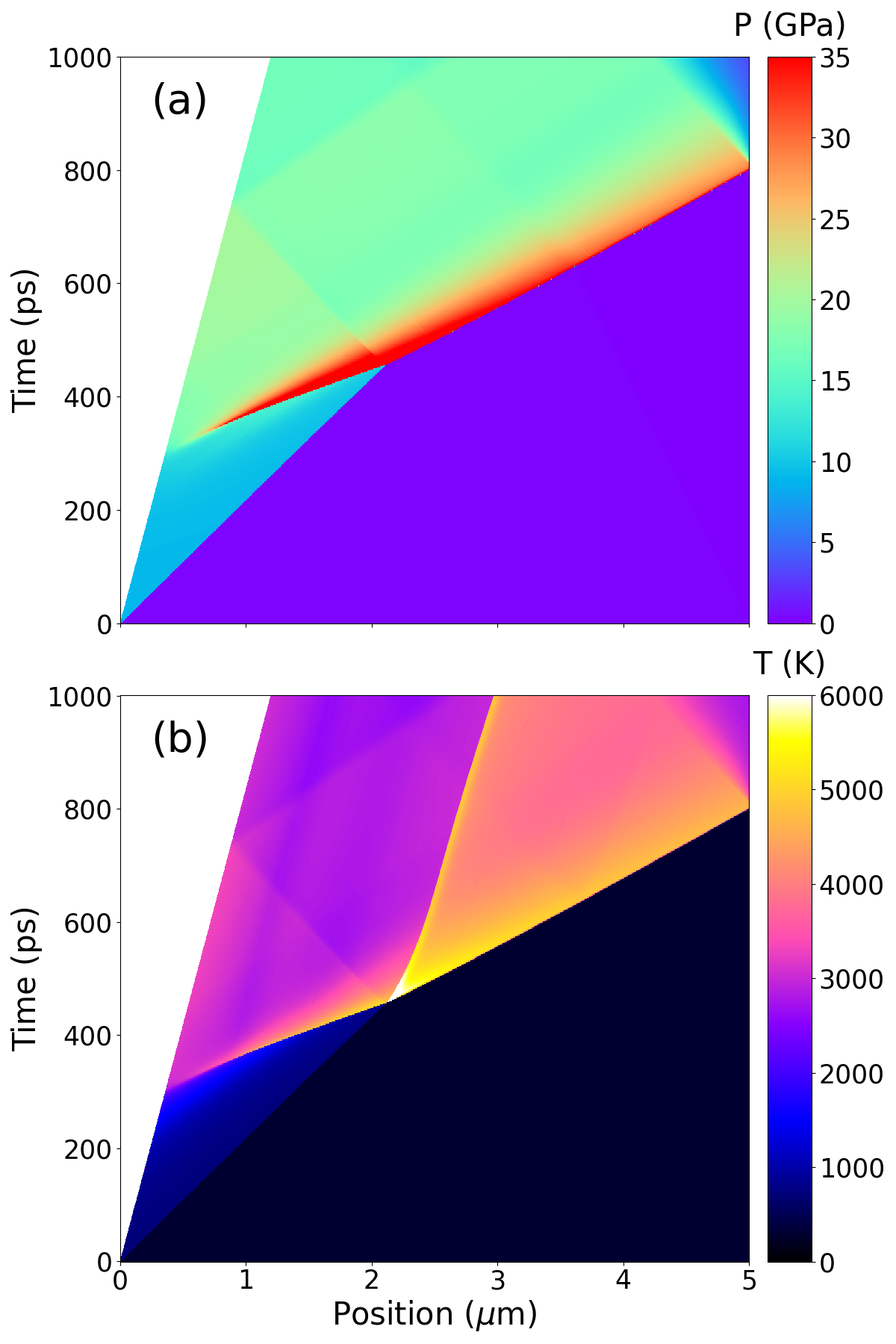}    
  \caption{Shock to detonation transition in single crystal $\beta$-HMX.}
  \label{fig:std}
\end{figure}

As a first application on the dynamic properties of $\beta$-HMX, the thermochemistry+EOS mesoscale model was used to study the response of a $\beta$-HMX single crystal to a sustained shock with piston velocity $U_p$ = 1200 m/s in 1D. A 5 $\mu$m sample was considered with 8192 nodes and a timestep of 20 fs. March diagrams for pressure and temperature are reported in Figure~\ref{fig:std}a and ~\ref{fig:std}b respectively.
The initial shock-wave heats the material accordingly to the inert EOS and a second abrupt increase in temperature is observed with a delay of approximately 150 ps, indicating the onset of a chemical decomposition of $\beta$-HMX. The slope of this overheating appears faster than the initial elastic wave announcing a transition to detonation which can be verified on the pressure march diagram where a second shock-wave is clearly visible. Eventually, the detonation wave reaches the initial elastic front. In addition, a slight pressure release wave can be observed going backwards after the reaction has initiated, which is attributed to the multi-components EOS. Investigating various conditions and simulation setups for the shock-to-detonation transition will be the object of future work and calculating ($\rho$,T,P) states using DFTB would help refining the multi-components EOS~\cite{Lozano_jap_2023}. 3D calculations involving macroscopic samples at the polycrystal scale will also be considered in order to compare to experiments, especially concerning the chemical reaction zone~\cite{Sollier_jap_2022}. The present work represents the first reactive MD fully informed mesoscale model for the shock-to-detonation transition and the present results are very promising. 

\section{5. Discussion}
\label{sec:discussion}
This paper presents a comprehensive approach to the multiscale modeling of HE, bridging the gap between MD simulations and FE methods. By incorporating specific deformation pathways and advanced analysis techniques, we have developed detailed constitutive models for TATB and HMX single crystals. These models include both mechanical and chemical aspects, providing a robust framework for simulating complex phenomena such as STD transition and hotspot criticality. 

Our findings demonstrate the potential of these multiscale models to enhance our understanding of the HE response under various conditions. In our opinion however, it is crucial to validate and refine such models through experimental comparisons. Future work will involve correlating simulation results with experimental data to ensure accuracy and reliability of the models. Laboratory scale experiments could be helpful to compare our MD-FE multiscale framework. In particular, nano-indentation at the single crystal scale or micropillar compression would help inform and adjust our mesoscale models. In addition, single crystal thermal ignition would allow for a specific validation of the thermochemistry model. Finally, the kinetics of our thermochemistry model should be compare to one dimensional time to explosion (ODTX) data and different existing models~\cite{Henson_tcc_2022,Manner_jacs_2024}. This iterative process of simulation and experimentation will help fine-tuning our approach, leading to more predictive and effective tools for the design and analysis of HEs.


\begin{thebibliography}{10}
\newcommand{\enquote}[1]{``#1''}

\bibitem{thompson_cpc_2022}
Thompson, A.~P., Aktulga, H.~M., Berger, R., Bolintineanu, D.~S., Brown, W.~M.,
  Crozier, P.~S., {in 't Veld}, P.~J., Kohlmeyer, A., Moore, S.~G., Nguyen,
  T.~D., Shan, R., Stevens, M.~J., Tranchida, J., Trott, C. and Plimpton,
  S.~J., \enquote{LAMMPS - a flexible simulation tool for particle-based
  materials modeling at the atomic, meso, and continuum scales,} \emph{Computer
  Physics Communications}, Vol. 271, p. 108171, 2022.

\bibitem{carrard_2024}
Carrard, T., Prat, R., Latu, G., Babilotte, K., Lafourcade, P., Amarsid, L. and
  Soulard, L., \enquote{ExaNBody: A HPC Framework for N-Body Applications,} in
  \enquote{Euro-Par 2023: Parallel Processing Workshops,} pp. 342--354,
  Springer Nature Switzerland, 2024.

\bibitem{lafourcade_prm_2019}
Lafourcade, P., Denoual, C. and Maillet, J.-B., \enquote{Mesoscopic
  constitutive law with nonlinear elasticity and phase transformation for the
  twinning-buckling of TATB under dynamic loading,} \emph{Phys. Rev.
  Materials}, Vol.~3, p. 053610, May 2019.

\bibitem{kroonblawd_cpc_2016}
Kroonblawd, M.~P., Mathew, N., Jiang, S. and Sewell, T.~D., \enquote{A
  generalized crystal-cutting method for modeling arbitrarily oriented crystals
  in 3D periodic simulation cells with applications to crystal–crystal
  interfaces,} \emph{Computer Physics Communications}, Vol. 207, pp. 232--242,
  2016.

\bibitem{negre_jpcm_2023}
Negre, C. F.~A., Alvarado, A., Singh, H., Finkelstein, J., Martinez, E. and
  Perriot, R., \enquote{A methodology to generate crystal-based molecular
  structures for atomistic simulations,} \emph{Journal of Physics: Condensed
  Matter}, Vol.~35, p. 225001, mar 2023.

\bibitem{guruprasad_asm_2023}
Guruprasad, T., Keryvin, V., Kermouche, G., Marthouret, Y. and Sao-Joao, S.,
  \enquote{Compressive behaviour of carbon fibres micropillars by in situ SEM
  nanocompression,} \emph{Composites Part A: Applied Science and
  Manufacturing}, Vol. 173, p. 107699, 2023.

\bibitem{franciosi_ijp_2015}
Franciosi, P., Le, L., Monnet, G., Kahloun, C. and Chavanne, M.-H.,
  \enquote{Investigation of slip system activity in iron at room temperature by
  SEM and AFM in-situ tensile and compression tests of iron single crystals,}
  \emph{International Journal of Plasticity}, Vol.~65, pp. 226--249, 2015.

\bibitem{lafourcade_jpcc_2018}
Lafourcade, P., Denoual, C. and Maillet, J.-B., \enquote{Irreversible
  Deformation Mechanisms for 1,3,5-Triamino-2,4,6-Trinitrobenzene Single
  Crystal through Molecular Dynamics Simulations,} \emph{The Journal of
  Physical Chemistry C}, Vol. 122, pp. 14954--14964, 2018.

\bibitem{ovito}
Stukowski, A., \enquote{{Visualization and analysis of atomistic simulation
  data with OVITO-the Open Visualization Tool},} \emph{Modelling and Simulation
  in Materials Science and Engineering}, Vol.~{18}, {1} {2010}.

\bibitem{freville_rsi_2023}
Fréville, R., Bruzy, N. and Dewaele, A., \enquote{{Optical full-field strain
  measurement within a diamond anvil cell},} \emph{Review of Scientific
  Instruments}, Vol.~94, p. 123905, 12 2023.

\bibitem{henry_am_2024}
Henry, L., Bruzy, N., Fréville, R., Denoual, C., Amadon, B., Églantine
  Boulard, King, A., Guignot, N. and Dewaele, A., \enquote{Martensitic-like
  microstructures across the isostructural phase transitions in Cerium,}
  \emph{Acta Materialia}, Vol. 271, p. 119863, 2024.

\bibitem{denoual_prl_2010}
Denoual, C., Caucci, A.~M., Soulard, L. and Pellegrini, Y.-P.,
  \enquote{Phase-Field Reaction-Pathway Kinetics of Martensitic Transformations
  in a Model ${\mathrm{Fe}}_{3}\mathrm{Ni}$ Alloy,} \emph{Phys. Rev. Lett.},
  Vol. 105, p. 035703, Jul 2010.

\bibitem{bruzy_jmps_2022}
Bruzy, N., Denoual, C. and Vattré, A., \enquote{Polyphase crystal plasticity
  for high strain rate: Application to twinning and retwinning in tantalum,}
  \emph{Journal of the Mechanics and Physics of Solids}, Vol. 166, p. 104921,
  2022.

\bibitem{lafourcade_jap_2024}
Lafourcade, P., Maillet, J.-B., Bruzy, N. and Denoual, C., \enquote{{Molecular
  dynamics informed calibration of crystal plasticity critical shear stresses
  for the mesoscopic mechanical modeling of
  1,3,5-triamino-2,4,6-trinitrobenzene (TATB) single crystal},} \emph{Journal
  of Applied Physics}, Vol. 135, p. 075901, 02 2024.

\bibitem{vattre_jmps_2016}
Vattré, A. and Denoual, C., \enquote{Polymorphism of iron at high pressure: A
  3D phase-field model for displacive transitions with finite elastoplastic
  deformations,} \emph{Journal of the Mechanics and Physics of Solids},
  Vol.~92, pp. 1--27, 2016.

\bibitem{trumel_europyro_2019}
Trumel, H., Rabette, F., Willot, F., Brenner, R., Ongari, E., Biessy, M. and
  Picart, D., \enquote{{Understanding the thermomechanical behavior of a
  TATB-based explosive via microstructure-level simulations. Part I:
  Microcracking and viscoelasticity},} in \enquote{{Europyro 44th International
  Pyrotechnics Seminar},} Tours, France, June 2019.

\bibitem{yeager_mat_2020}
Yeager, J.~D., Kuettner, L.~A., Duque, A.~L., Hill, L.~G. and Patterson, B.~M.,
  \enquote{Microcomputed X-Ray Tomographic Imaging and Image Processing for
  Microstructural Characterization of Explosives,} \emph{Materials}, Vol.~13,
  2020.

\bibitem{barrata2023dolfinx}
Barrata, I.~A., Dean, J.~P., Dokken, J.~S., Habera, M., Hale, J., Richardson,
  C., Rognes, M.~E., Scroggs, M.~W., Sime, N. and Wells, G.~N.,
  \enquote{DOLFINx: The next generation FEniCS problem solving environment,}
  2023.

\bibitem{alnaes2014unified}
Aln{\ae}s, M.~S., Logg, A., {\O}lgaard, K.~B., Rognes, M.~E. and Wells, G.~N.,
  \enquote{Unified form language: A domain-specific language for weak
  formulations of partial differential equations,} \emph{ACM Transactions on
  Mathematical Software (TOMS)}, Vol.~40, pp. 1--37, 2014.

\bibitem{van_duin_jpca_2001}
van Duin, A. C.~T., Dasgupta, S., Lorant, F. and Goddard, W.~A.,
  \enquote{ReaxFF: A Reactive Force Field for Hydrocarbons,} \emph{The Journal
  of Physical Chemistry A}, Vol. 105, pp. 9396--9409, 2001.

\bibitem{wood_prb_2018}
Wood, M.~A., Kittell, D.~E., Yarrington, C.~D. and Thompson, A.~P.,
  \enquote{Multiscale modeling of shock wave localization in porous energetic
  material,} \emph{Phys. Rev. B}, Vol.~97, p. 014109, Jan 2018.

\bibitem{yoo_npj_2021}
Yoo, P., Sakano, M., Desai, S., Islam, M., Liao, P. and Strachan, A.,
  \enquote{Neural network reactive force field for C, H, N, and O systems,}
  \emph{npj Comput. Mater.}, Vol.~7, 2021.

\bibitem{Cawkwell_jcp_2019}
Cawkwell, M.~J. and Perriot, R., \enquote{{Transferable density functional
  tight binding for carbon, hydrogen, nitrogen, and oxygen: Application to
  shock compression},} \emph{The Journal of Chemical Physics}, Vol. 150, p.
  024107, 01 2019.

\bibitem{sakano_jpca_2020}
Sakano, M.~N., Hamed, A., Kober, E.~M., Grilli, N., Hamilton, B.~W., Islam,
  M.~M., Koslowski, M. and Strachan, A., \enquote{Unsupervised Learning-Based
  Multiscale Model of Thermochemistry in 1,3,5-Trinitro-1,3,5-triazinane
  (RDX),} \emph{The Journal of Physical Chemistry A}, Vol. 124, pp. 9141--9155,
  2020, pMID: 33112131.

\bibitem{lafourcade_jpcc_2023}
Lafourcade, P., Maillet, J.-B., Roche, J., Sakano, M., Hamilton, B.~W. and
  Strachan, A., \enquote{Multiscale Reactive Model for
  1,3,5-Triamino-2,4,6-trinitrobenzene Inferred by Reactive MD Simulations and
  Unsupervised Learning,} \emph{The Journal of Physical Chemistry C}, Vol. 127,
  pp. 15556--15572, 2023.

\bibitem{fevotte_nc_2011}
Févotte, C. and Idier, J., \enquote{{Algorithms for Nonnegative Matrix
  Factorization with the $\beta$-Divergence},} \emph{Neural Computation},
  Vol.~23, pp. 2421--2456, 09 2011.

\bibitem{Lozano_jap_2022}
Lozano, E. and Aslam, T.~D., \enquote{{A robust three-parameter reference curve
  for condensed phase materials},} \emph{Journal of Applied Physics}, Vol. 131,
  p. 015902, 01 2022.

\bibitem{Lozano_jap_2023}
Lozano, E., Cawkwell, M.~J. and Aslam, T.~D., \enquote{{An analytic and
  complete equation of state for condensed phase materials},} \emph{Journal of
  Applied Physics}, Vol. 134, p. 125102, 09 2023.

\bibitem{handley_2012}
Handley, C.~A., \enquote{{Critical hotspots and flame propagation in HMX-based
  explosives},} \emph{AIP Conference Proceedings}, Vol. 1426, pp. 283--286, 03
  2012.

\bibitem{perriot_jap_2021}
Perriot, R. and Cawkwell, M.~J., \enquote{{Thermal conductivity tensor of
  $\beta$-1,3,5,7-tetranitro-1,3,5,7-tetrazoctane ($\beta$-HMX) as a function
  of pressure and temperature},} \emph{Journal of Applied Physics}, Vol. 130,
  p. 145106, 10 2021.

\bibitem{tarver_1996}
Tarver, C.~M., Chidester, S.~K. and Nichols, A.~L., \enquote{Critical
  Conditions for Impact- and Shock-Induced Hot Spots in Solid Explosives,}
  \emph{The Journal of Physical Chemistry}, Vol. 100, pp. 5794--5799, 1996.

\bibitem{hamilton_jpca_2021}
Hamilton, B.~W., Steele, B.~A., Sakano, M.~N., Kroonblawd, M.~P., Kuo, I.-F.~W.
  and Strachan, A., \enquote{Predicted reaction mechanisms, product speciation,
  kinetics, and detonation properties of the insensitive explosive 2,
  6-diamino-3, 5-dinitropyrazine-1-oxide (llm-105),} \emph{The Journal of
  Physical Chemistry A}, Vol. 125, pp. 1766--1777, 2021.

\bibitem{Sollier_jap_2022}
Sollier, A., Hébert, P. and Letremy, R., \enquote{{Chemical reaction zone
  measurements in pressed trinitrotoluene (TNT) and comparison with
  triaminotrinitrobenzene (TATB)},} \emph{Journal of Applied Physics}, Vol.
  131, p. 055902, 02 2022.

\bibitem{Henson_tcc_2022}
Henson, B. and Smilowitz, L., \enquote{Chapter 15 - Chemical kinetics and the
  decomposition of secondary explosives,} in \enquote{Molecular Modeling of the
  Sensitivities of Energetic Materials,} , edited by Mathieu, D., Vol.~22 of
  \emph{Theoretical and Computational Chemistry}, pp. 369--402, Elsevier, 2022.

\bibitem{Manner_jacs_2024}
Manner, V.~W., Cawkwell, M.~J., Spielvogel, K.~D., Tasker, D.~G., Rose, J.~W.,
  Aloi, M., Tucker, R., Moore, J.~D., Campbell, M.~C. and Aslam, T.~D.,
  \enquote{An Integrated Experimental and Modeling Approach for Assessing
  High-Temperature Decomposition Kinetics of Explosives,} \emph{Journal of the
  American Chemical Society}, Vol. 146, pp. 26286--26296, 2024, pMID: 39259775.

\end{thebibliography}

\end{document}